\documentclass[a4paper,11pt]{article}
\pdfoutput=1 
\usepackage{jheppub} 
\usepackage{color}
\usepackage{graphicx, multirow,soul,amsmath,amsfonts,amssymb,mathrsfs,amsfonts}
\usepackage{hyperref}
\usepackage[all]{hypcap}
\usepackage{url}
\usepackage{bbm}
\usepackage{cancel}
\usepackage{slashed}
\usepackage[dvipsnames]{xcolor}
\usepackage{multicol, blindtext}
\usepackage{lipsum}
\usepackage{mattens}
\usepackage{mathtools}
\usepackage{footnote}
\usepackage{subcaption}
\usepackage[utf8]{inputenc}

\newcommand{\pythia}[1]{\textsc{Pythia#1}}

\newcommand{\madgraph}[1]{\textsc{MadGraph#1}}

\newcommand{\feynrules}[1]{\textsc{FeynRules#1}}

\newcommand{\delphes}[1]{\textsc{Delphes#1}}

\newcommand{\met}{\slashed{E}_{\mathrm{T}}}
\newcommand{\keV}{\ensuremath{\rm~keV}\xspace}

\newcommand{\GeV}{\ensuremath{\rm~GeV}\xspace}
\newcommand{\TeV}{\ensuremath{\rm~TeV}\xspace}

\newcommand{\be}{\begin{equation}}
\newcommand{\ee}{\end{equation}}
\newcommand{\iee}{{\it i.e.}}
\newcommand{\egg}{{\it e.g.}}
\newcommand{\msq}{\ensuremath{M_{\tilde{q}}}}
\newcommand{\mbino}{\ensuremath{M_{\tilde{B}}}}
\newcommand{\octino}{\mathcal{O}}

\preprint{\\ UCI-HEP-TR-2018-21\\IFT-UAM/CSIC-18-122 
\\FTUAM-18-27\\  FERMILAB-PUB-19-010-T}

\title{Bi$\boldsymbol{\nu}$o phenomenology at the LHC}

\author[a, b]{Julia Gehrlein,}
\author[c]{Seyda Ipek,}
\author[d]{and Patrick J. Fox}

\affiliation[a]{Instituto de F\'isica Te\'orica UAM/CSIC, Calle Nicol\'as Cabrera 13-15, Cantoblanco E-28049 Madrid, Spain}
\affiliation[b]{Departamento  de  F\'{\i}sica Te\'{o}rica,  Universidad  Aut\'{o}noma  de  Madrid, Cantoblanco  E-28049  Madrid,  Spain}
\affiliation[c]{Department of Physics and Astronomy, University of California, Irvine
4129 Frederick Reines Hall, Irvine, CA 92617-4575, U.S.A.}
\affiliation[d]{Theoretical Physics Department, Fermi National Accelerator Laboratory, P.O. Box 500, Batavia, IL 60510, USA.}

\abstract{We study the LHC constraints on an $R$-symmetric SUSY model, where the neutrino masses are generated through higher dimensional operators involving the pseudo-Dirac bino, named bi$\nu$o. We consider a particle spectrum where the squarks are heavier than the lightest neutralino, which is a pure bi$\nu$o.  The bi$\nu$o is produced through squark decays and it subsequently decays to a combination of jets and leptons, with or without missing energy, via its mixing with the Standard Model neutrinos. We recast the most recent LHC searches for  jets+$\met$ with $\sqrt{s}=13~$TeV and $\mathcal{L}=36~{\rm fb}^{-1}$ of data to determine the constraints on the squark and bi$\nu$o masses in this model.  
We find that squarks as light as 350~GeV are allowed if the bi$\nu$o is lighter than 150~GeV and squarks heavier than 950~GeV are allowed for any bi$\nu$o mass. We also present forecasts for the LHC with $\sqrt{s}=13$~TeV and $\mathcal{L}=300~{\rm fb}^{-1}$ and show that squarks up to 1150~GeV can be probed.
}

\begin{document} 
\maketitle
\flushbottom

%%%%%%%%%%%%%%%%%
\section{Introduction}
\label{sec:intro}
The Standard Model (SM) can explain a multitude of observations. However, several phenomena still require explanations, \egg~the existence and nature of dark matter, the matter--antimatter asymmetry of the universe, and the origin of neutrino masses. 

A popular model to explain these beyond the SM physics is minimal supersymmetry (MSSM). Although the MSSM addresses issues of fine-tuning in the Higgs mass and there are dark matter candidates in MSSM, it has been constrained stringently by LHC searches \cite{CMSsusy, ATLASsusy}.  The MSSM currently also lacks a simple mechanism to generate neutrino masses as well as the baryon asymmetry of the universe. As such, it is necessary to consider supersymmetric models beyond the MSSM. 

One extension of the MSSM that addresses these questions is the $R$-symmetric MSSM \cite{HALL1991289}. In $R$-symmetric MSSM the superpartners are charged under a global $U(1)_R$ symmetry while their SM counterparts are neutral.  While this global symmetry is unbroken, gauginos cannot be Majorana particles. Additional adjoint fields with opposite $U(1)_R$ charge, with respect to gauginos, are introduced so that gauginos can acquire Dirac masses \cite{FAYET1975104, FAYET1976135}. $R$-symmetric MSSM addresses SUSY $CP$ and flavor problems by forbidding one-loop diagrams mediated by Majorana gauginos as well as forbidding left-right sfermion mixing \cite{Kribs:2007ac, Dudas:2013gga}. Apart from gauginos, this model requires electroweak (EW) partners for higgsinos with an opposite $R$-charge so that a $\mu$-term is allowed. It was shown that the scalar components of these new superfields can help to have a first-order EW phase transition \cite{Fok:2012fb}. Moreover, new interactions can bring in new sources of \emph{CP} violation. Hence this model can potentially explain the baryon asymmetry of the universe\footnote{Another mechanism for generating the baryon asymmetry in such models is oscillations and out-of-equilibrium decays of a pseudo-Dirac bino \cite{Ipek:2016bpf}.} \cite{Fok:2012fb}. Furthermore, Dirac gluinos make the fine-tuning problem milder in $R$-symmetric MSSM \cite{Kribs:2007ac, Fox:2002bu}. 

The global  $U(1)_R$ symmetry is broken because the gravitino acquires a mass. Consequently small $U(1)_R$-breaking Majorana masses for gauginos will be generated through anomaly mediation \cite{Randall:1998uk, Giudice:1998xp, ArkaniHamed:2004yi}. Since the $U(1)_R$ symmetry is only approximate, gauginos in $R$-symmetric MSSM are pseduo-Dirac fermions, having both Dirac and Majorana masses.  

The LHC phenomenology of $R$-symmetric MSSM is different than minimal SUSY models. For example in $R$-symmetric MSSM the supersymmetric particles need to be produced in particle--antiparticle pairs since the initial SM state is $U(1)_R$ symmetric. Furthermore some production channels for supersymmetric particles are not available due to the $U(1)_R$ symmetry. Hence collider limits on $R$-symmetric MSSM tend to be less stringent than the ones on MSSM, see \emph{e.g.}, \cite{Frugiuele:2012kp, Alvarado:2018rfl,Diessner:2017ske,Kalinowski:2015eca}. 

In this work we study the LHC phenomenology of a version of the $R$-symmetric MSSM in which the $U(1)_R$ symmetry is elevated to $U(1)_{R-L}$, where $L$ is the lepton number. We give the details of the model in Section~\ref{sec:model}.  It has been shown that in this model the pseudo-Dirac bino can play the role of right-handed neutrinos \cite{Coloma:2016vod} and light Majorana neutrino masses are generated via an inverse-seesaw mechanism. The smallness of the light neutrino masses is given by a hierarchy between the source of $U(1)_R$-breaking, namely the gravitino mass $m_{3/2}$, and the messenger scale $\Lambda_M$.  As benchmark points this requires $m_{3/2}\sim$~10~keV and $\Lambda_M\sim$~100~TeV. 

The mixing between electroweak gauginos and the SM neutrinos allows the gauginos to decay to gauge bosons and leptons, which can remove the usual $\met$ signature associated with SUSY searches.  In cases where the lepton is a neutrino there is still $\met$ in the event but the kinematics are different from typical weak scale SUSY models.
We use current searches for jets+$\met$ at the LHC with $\sqrt{s}=13~$TeV and  $\mathcal{L}=36~{\rm fb}^{-1}$ to find the constraints on squark and bino masses in this model. We focus on the parameter region with $100~{\rm GeV}<\mbino<\msq$. We also forecast our results for $\sqrt{s}=13~$TeV and $\mathcal{L}=300~{\rm fb}^{-1}$. The analysis is described in Section~\ref{sec:LHCpheno}. Our results are shown in  Figure~\ref{fig:Plotexcatlas} and our conclusions are given in Section~\ref{sec:conclusions}.

%%%%%%%%%%%%%%%%%%%%%%%%%%%
\section{Model}
\label{sec:model}

In this section we review the model that was considered in \cite{Coloma:2016vod}. This is an extension of $U(1)_R$--symmetric SUSY models \cite{Kribs:2007ac} where, instead of  the $R$ symmetry, the model has a global $U(1)_{R-L}$ symmetry. The field content and the $U(1)_R$ and $U(1)_{R-L}$ charges of the relevant superfields are given in Table \ref{table:fields}. Note that in the rest of the text we use $U(1)_R$ instead of $U(1)_{R-L}$ whenever the distinction is not important.

$U(1)_R$--symmetric SUSY is an extension of the MSSM in which the superpartners of the SM particles are charged under a global $U(1)$ symmetry. The SM particles are not charged under this symmetry.  This model was introduced \cite{Kribs:2007ac} to solve the SUSY \emph{CP} and flavor problems. Due to the $U(1)_R$ symmetry, Majorana masses for the gauginos are forbidden as well as left-right mixing of sfermions. Hence, \emph{e.g.}, one-loop diagrams that would generate a large electric dipole moment for fermions are suppressed, solving the SUSY \emph{CP} problem. Similar arguments follow for the flavor problem.

\begin{table}[t]
\centering
\begin{tabular}{|c|c|c|}
\hline
Superfields	&	$U(1)_R$	&	$U(1)_{R-L}$ \\
\hline\hline
$Q, U^c, D^c$	&	 1	&	1 \\
$L$	&	1	&	0 \\
$E^c$	&	1	&	2	\\	
\hline
$H_{u,d}$	&	0	&	0	\\
$R_{u,d}$	&	2	&	2	\\
\hline
$W_{\tilde{B},\tilde{W},\tilde{g}}$	&	1	&	1	\\
$\Phi_{S,T,\octino}$	&	0	&	0	\\
\hline
gravitino/goldstini & 1 &  1 \\ 
\hline
\end{tabular}
\caption{The relevant field content of the model. (SM charges are not shown.) $\Phi_{S,T,\octino}$ are superfields which has the same SM charges as $W_{\tilde{B},\tilde{W},\tilde{g}}$ and their fermionic components, $S, T,\octino$ are the Dirac partners of the bino, wino and the gluino respectively. The fermionic components of the superfields $R_{u,d}$ are the Dirac partners of the Higgsinos $\tilde{h}_{u,d}$.} \label{table:fields}
\end{table}

Phenomenologically one novel aspect of $U(1)_R$--symmetric SUSY is that gauginos in this model are not Majorana fermions, since they are charged under a global symmetry, but are instead Dirac particles. In order to make gauginos into Dirac fermions, adjoint fields are added for each SM gauge field with opposite $U(1)_R$ charges \cite{Fox:2002bu}. These new fields, $\Phi_{S,T,\octino}$ are called singlino, tripletino and octino respectively and their fermionic components, $S,T,\octino$, become the Dirac partners of the bino, weakino and the gluino\footnote{Dirac gauginos have been studied in the literature extensively. See, \emph{e.g.}, \cite{Benakli:2010gi, Benakli:2008pg, Goodsell:2012fm}}.  The Dirac nature of gauginos means that $t$-channel gluino exchange diagrams which contribute to squark pair production are suppressed and the rate for squark production at the LHC is reduced, allowing for lighter squarks.  Such models have been dubbed ``super-safe" \cite{Kribs:2012gx}.  Furthermore, the minimal incarnation of Dirac gauginos, supersoft SUSY breaking \cite{Fox:2002bu}, has only a $D$-term spurion leading to improved renormalization properties.  Sfermions only receive finite contributions to their mass, rather than logarithmically divergent as in $F$-term breaking scenarios.  The ratio between gluino and squark masses is also larger, $m_{\tilde{g}}/m_{\tilde{f}}\sim 5-10$, than in alternative scenarios. 

The unbroken $R$-symmetry forbids the usual $\mu$-term.  In order to give mass to higgsinos, superfields $R_{u,d}$, with the same SM charge as the Higgs superfield but opposite $U(1)_R$ charges are added.  While the usual two Higgs doublets $H_{u,d}$ acquire vacuum expectation values (vev), $R_{u,d}$ do not.

Here we will be considering several sources of SUSY breaking, both $F$- and $D$-term.  We envision two sectors each of which separately break SUSY. The first contains both a $D$-term and $F$-term spurion, of comparable size, and is coupled to the fields in the supersymmetric standard model.  The second is not coupled to the standard model, except through gravity and has a higher SUSY breaking scale than the first sector.  We are agnostic as to whether this is in $F$- or $D$-terms, or both, and parametrize the SUSY breaking simply as $F_2$. This second sector will raise the mass of the gravitino ($m_{3/2}$) and provide an additional goldsitino with tree-level mass $2 m_{3/2}$ \cite{Cheung:2010mc,Cheung:2011jq}.

%%%%%%%%
\subsection{SUSY breaking and superpartner masses}

We focus for now on the effects of the SUSY breaking that is communicated non-gravitationally to the SM.  SUSY is broken in a hidden sector which communicates with the visible sector at the messenger scale $\Lambda_M$. SUSY breaking is incorporated via the spurions,
\begin{align}
W'_\alpha = \theta_\alpha D~,\qquad X=\theta^2 F~.
\end{align}
We assume that $X$ transforms non-trivially under some symmetry of the SUSY-breaking sector so that gauginos do not have Majorana masses of the form $\int d^2\theta (X/\Lambda_M)W_\alpha W^\alpha$, where $W^\alpha$ is a SM gauge field strength superfield. $W'_\alpha$ is the field strength of a hidden $U(1)'$ which gets a $D$-term vev. In this case, Dirac gaugino masses come from the supersoft term \cite{Fox:2002bu}
\begin{align}\label{eq:supersoftop}
\int d^2\theta\, \frac{\sqrt{2} c_i}{\Lambda_M} W'_\alpha W_i^\alpha\Phi_i~,
\end{align}
where $c_i$ is a dimensionless coefficient, that we take to be $\mathcal{O}(1)$, and $i=\tilde{B}, \tilde{W}, \tilde{g}$. This operator can be generated by integrating out messenger fields, of mass $\sim \Lambda_M$,  charged under both the SM and the $U(1)'$. The Dirac mass of the gaugino is $M_i=c_i D/\Lambda_M$.  
The operator (\ref{eq:supersoftop}) also gives a mass to the scalar adjoint, while leaving the pseudoscalar massless\footnote{Pseudoscalars in the extended superpartners can acquire masses through another soft term of the form $\int d^2\theta \frac{W'_\alpha W^{'\alpha}}{\Lambda_M^2}\Phi_i^2$~\cite{Fox:2002bu}.}, and introduces a trilinear coupling between the scalar adjoint, the SM and the $D$-term.  At one loop a scalar charged under gauge group $i$ receives a \emph{finite} soft mass from the gaugino
\be
m^2 = \frac{C_i \alpha_i \left(M_i\right)^2}{\pi}\log 4~,
\ee
where $C_i$ is the quadratic Casimir of the scalar and we have assumed the scalar adjoint only receives a mass from (\ref{eq:supersoftop}). We will be interested in a spectrum with the bino in the $\mathcal{O}(100~{\rm GeV} - {\rm TeV})$ mass range and the squarks in the same range, but heavier than the bino.  If the sfermion masses are entirely from the supersoft operator this means the right-handed sleptons would be below the LEP bound.  Thus, at least for the right handed sleptons, we include additional sources of SUSY breaking through the operator
\be
\int d^4\theta\, \frac{X^\dagger X}{\Lambda_M^2}c_{ij} \Psi^\dagger_i \Psi_j~,
\ee 
with $\Psi_i$ a right handed lepton superfield.  We assume $F\sim D$ and $c_{ij}\sim 1$.
The squarks can be heavier than the bino from the finite supersoft contributions alone, as long as the gluino is sufficiently heavy, in the multi-TeV mass range.

As all global symmetries, $U(1)_R$ is broken due to gravity. Anomaly mediation \cite{Randall:1998uk} generates a Majorana mass for the gauginos proportional to the gravitino mass, $m_{3/2}$, 
\begin{align}
m_i=\frac{\beta(g)}{g}m_{3/2}~,
\end{align}
where $\beta(g)$ is the beta function for the appropriate SM gauge coupling $g$.  The gravitino picks up mass from all sources of SUSY breaking, $m_{3/2}^2 = \sum_i (F_i^2 + D_i^2/2)/\sqrt{3} M_{\rm Pl}^2$.  We assume that the messenger scale $\Lambda_M$ is below the Planck scale and thus $m_i\ll M_i$.  We ignore the small anomaly mediated corrections to scalar masses.
Note that $U(1)_R$-breaking Majorana masses for the Dirac partners, $\tilde{m}_i \Phi_i \Phi_i$, could also be generated. We assume these are much smaller than the Dirac gaugino masses as well. (For LHC studies we will set the Majorana masses to zero.)
Due to the small anomaly-mediated Majorana gaugino masses, the gauginos are pseudo-Dirac particles.

%%%%%%%%%
\subsection{Neutrino masses}
\label{sec:neutrinomasses}

It has been shown in \cite{Coloma:2016vod} that the operators,
\be
\frac{f_i}{\Lambda_M^2}\int d^2\theta\, W'_\alpha W_{\tilde{B}}^\alpha H_u L_i \ \ \ \text{and}\ \ \ \frac{d_i}{\Lambda_M}\int d^4\theta\,  \phi^\dagger \Phi_S H_u L_i
\ee
(where $\phi=1+\theta^2 m_{3/2}$) can generate two non-zero neutrino masses through the Inverse Seesaw mechanism
\cite{Mohapatra:1986aw,Mohapatra:1986bd}, with the bino--singlino pair acting as a pseudo-Dirac right-handed neutrino.  These operators can be generated by integrating out two pairs of gauge singlets $N_i,N_i'$, with R-charge 1 and lepton number $\mp 1$.

Once the Higgs acquires a vev the neutrino-bino mass mixing matrix, in the basis $(\nu_i, \tilde{B}, S)$, is 
\be
\mathbb{M} =	\begin{pmatrix}
			0_{3\times3}	&	\mathbf{Y}v	&	\mathbf{G}v \\
			\mathbf{Y}^Tv	&	m_{\tilde{B}}	&	M_{\tilde{B}}		\\
			\mathbf{G}^Tv	&	M_{\tilde{B}}			&	m_S
			\end{pmatrix}~, \label{eq:Mneutrino}
\ee			
with $Y_i= f_i M_{\tilde{B}}/\Lambda_M$ and $G_i = d_i m_{3/2}/\Lambda_M$.  The mass matrix $\mathbb{M}$ has an Inverse Seesaw structure with $\mathbf{G}\ll \mathbf{Y}$.  The light neutrino masses do not depend on the Dirac bino mass and at normal ordering they are given by
\begin{align}
m_1 = 0,~~~m_2=\frac{m_{3/2}\, v^2}{\Lambda_M^2}(1-\rho),~~~ m_3=\frac{m_{3/2}\, v^2}{\Lambda_M^2}(1+\rho)~,
\end{align}
where $\rho = \hat{\mathbf{Y}}\cdot\hat{\mathbf{G}}$,  which is determined by the neutrino mass splittings to be $\simeq0.7$. We ignore the small corrections, $\mathcal{O}(m_S/M_D)$, due to Majorana masses.
Parametrically the neutrino masses are
 \begin{align}
 m_\nu\simeq (2-20)\times10^{-2}~{\rm eV}\left(\frac{m_{3/2}}{10~{\rm keV}}\right)\left(\frac{100~{\rm TeV}}{\Lambda_M}\right)^2~.
 \end{align}
To recover the correct neutrino mixing matrix, and by setting all phases in the neutrino sector to zero, $\mathbf{Y}$ and $\mathbf{G}$ must have the approximate form
\be
\mathbf{Y}\simeq \frac{M_{\tilde{B}}}{\Lambda_M}
\begin{pmatrix}
0.35 \\
0.85 \\
0.35
\end{pmatrix},\quad
\mathbf{G}\simeq \frac{m_{3/2}}{\Lambda_M}
\begin{pmatrix}
-0.06 \\
0.44 \\
0.89
\end{pmatrix}~. \label{eq:YG}
\ee

Low-energy searches for lepton flavor violation place strong constraints on these couplings. The strongest current constraint comes from ${\rm Br}(\mu\to e\gamma)$ \cite{TheMEG:2016wtm} and places a lower bound on the messenger scale $\Lambda_M>30$~TeV, independent of $M_{\tilde{B}}$ or $m_{3/2}$.  Future experiments, \emph{e.g.} Mu2e \cite{Bartoszek:2014mya}, will probe $\Lambda_M\sim100$~TeV. We use the word ``bi$\nu$o" from now on to refer to the pseudo-Dirac bino in order to emphasize that it is involved in neutrino-mass generation.

%%%%%%%%%%
\subsection{Neutralino mixing}

In $R$-symmetric models the Higgs sector is extended by two additional $SU(2)$ doublets, $R_{u,d}$, that do not acquire a vev.  Once electroweak symmetry is broken there is mixing between $R_{u,d}$ and the adjoint fermions, $S$ and $T$, in addition to the usual wino-bi$\nu$o mixing.  However, the neutrinos only mix with the bi$\nu$o.  Significant neutralino mixing only changes the collider phenomenology and does not affect the generation of neutrino masses, which happens only through bi$\nu$o--neutrino mixing.  We follow \cite{Kribs:2008hq} to investigate the neutralino mixing in this model.

The relevant part of the superpotential for neutralino mixing is 
\begin{align}\label{eq:Wneutralino}
\mathcal{W}=\mu_u H_u R_u + \mu_d H_d R_d + \Phi_S \left( \lambda^u_{\tilde{B}} H_u R_u + \lambda^d_{\tilde{B}} H_d R_d  \right) + \Phi_T \left( \lambda^u_{\tilde{W}} H_u R_u + \lambda^d_{\tilde{W}} H_d R_d  \right).
\end{align}
After EW symmetry breaking, together with Dirac gaugino masses, kinetic terms and ignoring the small Majorana gaugino masses, (\ref{eq:Wneutralino}) generates the neutralino mass matrix 
\begin{align} \label{eq:Mneutralino}
\mathbb{M}_N=\begin{pmatrix}
		M_{\tilde{B}}	&	0	&	\frac{g_Y v_u}{\sqrt{2}}	&	-\frac{g_Y v_d}{\sqrt{2}} \\
		0			&	M_{\tilde{W}}	&	-\frac{g_2 v_u}{\sqrt{2}}	&	\frac{g_2 v_d}{\sqrt{2}} \\
		\frac{\lambda_{\tilde{B}}^u v_u}{\sqrt{2}}	&	-\frac{\lambda_{\tilde{W}}^u v_u}{\sqrt{2}}	&	\mu_u	&	0	\\
		-\frac{\lambda_{\tilde{B}}^d v_d}{\sqrt{2}}	&	\frac{\lambda_{\tilde{W}}^d v_d}{\sqrt{2}}	&	0	&	\mu_d
\end{pmatrix}
\end{align}
in the basis $( \tilde{B},\tilde{W}, \tilde{R}_u, \tilde{R}_d )\times ( S,T,\tilde{h}_u, \tilde{h}_d )$, where $\tilde{R}_{u,d}$ are the fermionic components of the superfield $R_{u,d}$ (see Table \ref{table:fields}). Here $v_{u,d}\equiv \langle H_{u,d}\rangle$ are the up/down-type Higgs vevs defined as $v_u^2+v_d^2=v^2/2 \simeq(174~{\rm GeV})^2$ and $M_{\tilde{B},\tilde{W}}$ are the bi$\nu$o and wino Dirac masses defined in (\ref{eq:supersoftop}).

The neutralino mass matrix $\mathbb{M}_N$ has a rather simple form due to the Dirac nature of gauginos. It further simplifies for large $\tan\beta \equiv v_u/v_d$. In this limit
\begin{align}
\mathbb{M}_N\simeq\begin{pmatrix}
		M_{\tilde{B}}	&	0	&	\frac{g_Y v}{2}	&	0 \\
		0			&	M_{\tilde{W}}	&	-\frac{g_2 v}{\sqrt{2}}		&	0 \\
		\frac{\lambda_{\tilde{B}}^u v}{2}	&	-\frac{\lambda_{\tilde{W}}^u v}{2}	&	\mu_u	&	0	\\
		0	&	0	&	0	&	\mu_d
\end{pmatrix}.
\end{align}
It can immediately be seen that one of the states, with mass $\mu_d$, decouples. Furthermore, in the limit where $\lambda_{\tilde{B}}^u = \lambda_{\tilde{W}}^u=0$, there is no mixing between the bi$\nu$o, weakino and the Higgsinos. For simpicity, we assume a hierarchy $\mu>M_{\tilde{B},\tilde{W}}$ and work in this limit, where the lightest neutralino is a pure bi$\nu$o. 

\begin{figure}[t]
\centering
\includegraphics[width=.7\textwidth]{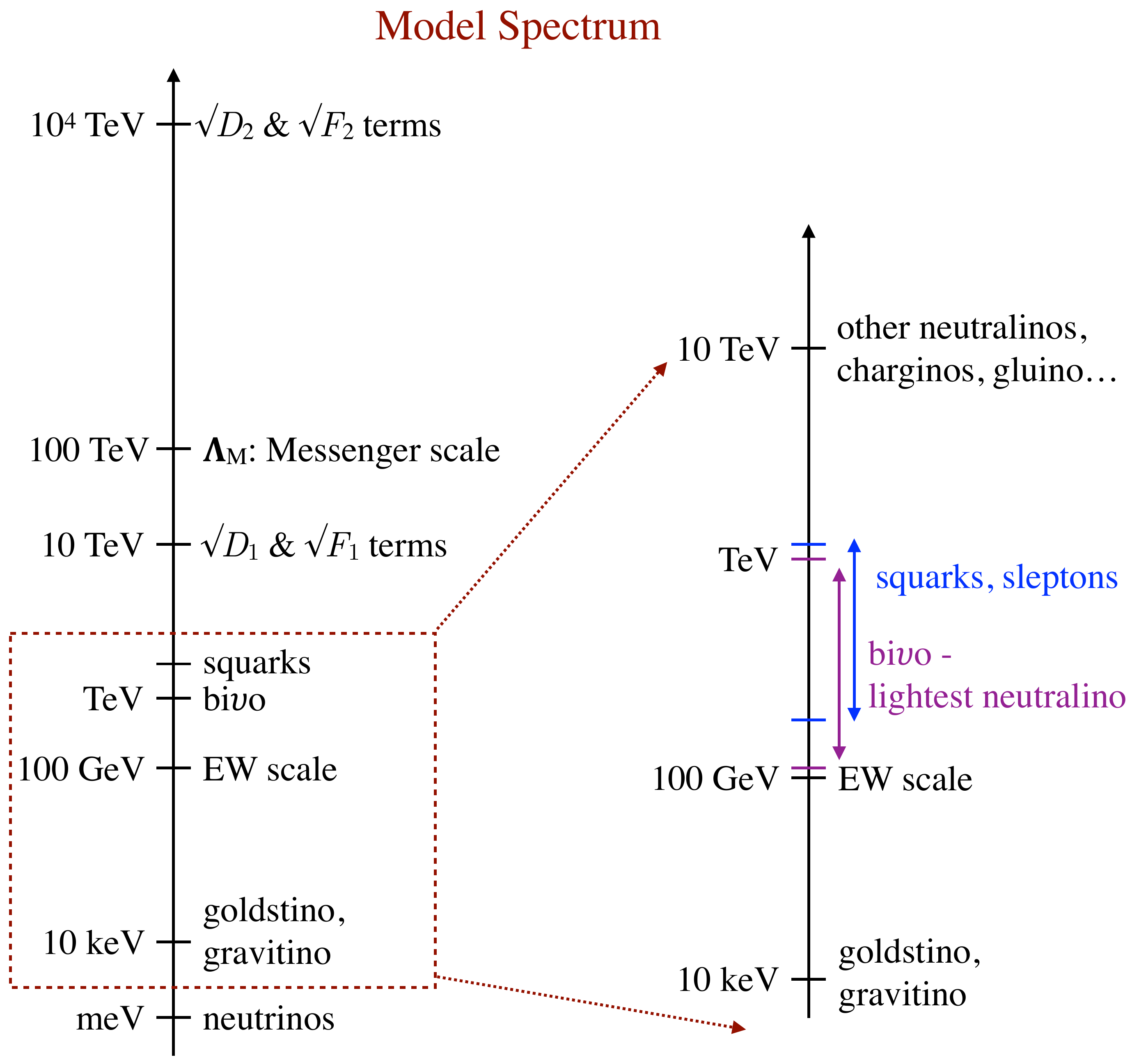}
\caption{Approximate spectrum of particles in the model described in Section \ref{sec:model}.}\label{fig:spectrum}
\end{figure}

%%%%%%%%%%%
\subsection{Gravitino/Goldstino dark matter}
\label{sec:goldstini}

As discussed in Section~\ref{sec:model}, the model has two independent sectors that break supersymmetry.  The breaking at the lower scale involves a $D$-term spurion but for the purposes of the discussion here it is sufficient to parametrize the two breaking scales as $\tilde{F}_{(1,2)}$, with $\tilde{F}^2=F^2+D^2/2$ and $\tilde{F}_{2}>\tilde{F}_{1}$.  A viable neutrino mass spectrum, and the spectrum of superparticles we are interested in, is achieved with $F_1\sim (10\TeV)^2$ and $F_2\sim  (10^4\TeV)^2$, as shown in Figure~\ref{fig:spectrum}.

Since there are two independent sources of SUSY breaking, there are two goldstini \cite{Cheung:2010mc,Cheung:2011jq}, of which one linear combination is eaten by the gravitino to have mass $m_{3/2} = \sqrt{(\tilde{F}_{1})^2 + (\tilde{F}_{2})^2}/\sqrt{3}M_{\mathrm{Pl}}\sim 10\keV$, while the other is twice as heavy, at tree level.  Furthermore, the couplings of the uneaten goldstini are enhanced relative to the gravitino's by a factor of $\tilde{F}_2/\tilde{F}_1$.  

Both the gravitino, $\tilde{G}$, and the golstino, $\zeta$, are lighter than the other $R$-symmetry-odd particles.  The goldstino can decay into a gravitino and SM particles, \emph{e.g.} $\zeta\rightarrow \tilde{G} \psi\bar{\psi}$.  The lifetime for this process is 
\be
\tau_{\zeta\rightarrow \tilde{G}\psi\bar{\psi}} \sim \frac{9\pi^3}{4} \frac{M_{\mathrm{Pl}}^4}{m_{3/2}^5}\left(\frac{\tilde{F}_1}{\tilde{F}_2}\right)^2~.
\ee
Furthermore, even though the gravitino in this model is the LSP, it can decay into neutrinos and photons via the neutrino-bi$\nu$o mixing. The gravitino lifetime is $\tau = \Gamma^{-1} \sim \frac{M_{\rm Pl}^2}{\theta^2 m_{3/2}^3}\sim 10^{39}$~s for $m_{3/2}\sim10$~keV and the bi$\nu$o-neutrino mixing angle $\theta \sim Y v/M_{\tilde{B}}\sim 10^{-3}$. Thus, for the range of parameters we are interested in, both the gravitino and goldstino are cosmologically stable and may contribute to dark matter.

It has been shown that a gravitino with mass $O(1-10~{\rm keV})$ could be a warm dark matter candidate \cite{Takayama:2000uz, Gorbunov:2008ui, Cheung:2011nn, Monteux:2015qqa}. The parameter region we study in this model suggests that gravitino could be a dark matter candidate if $T_{\rm reh} \sim O({\rm TeV})$. However, with the same parameters, goldstino would be overproduced since its couplings are enhanced by a factor of $\tilde{F}_2/\tilde{F}_1$. The abundance of goldstino depends on the production mechanism and $T_{\rm reh}$. (Depending on the masses of the gravitino, goldstino and other sparticles, the dominant production channel can be either decays or scatterings.) If $T_{\rm reh}< M_{\tilde{B},\tilde{q}}$, the sparticle abundance, and hence the abundance of goldstinos, will be suppressed. One expects a range of reheat temperatures where there will be just enough goldstino/gravitino to make up the correct dark matter abundance. Finding this range requires detailed calculations for allowed range of sparticle masses. We leave this for future work.

%%%%%%%%%%%%%%%%%%%%%%%%%%%%%
\section{LHC phenomenology}\label{sec:LHCpheno}

In this section we recast current LHC searches to find the constraints on the model described in Section \ref{sec:model}. In order to make the LHC analysis more tractable, we assume the following mass hierarchy for the SUSY particles (see Figure \ref{fig:spectrum}). 
\begin{itemize}
\item Gravitino is the LSP with $m_{3/2}\sim O(10~ {\rm keV})$.
\item Next-to-lightest supersymmetric particle (NLSP) is a pure bi$\nu$o, and the other neutralinos are decoupled. Note that there are two physical bi$\nu$o states with masses $M_{\tilde{B}}\pm \frac{m_{\tilde{B}}+m_S}{2}$. For simplicity we take the Majorana masses to be zero in the LHC analysis. Hence the physical bi$\nu$o mass is $M_{\tilde{B}}$. 
\item Squarks are degenerate and heavier than the bi$\nu$o. We do not apply any flavor tags in the analyses and only consider the first two generations of squarks, which gives a conservative estimate for the rate. 
\item Slepton masses are of the same order as squarks, and slepton production is irrelevant.
\item As expected for an $R$-symmetric model, the gluino and charginos are considerably heavier than the sfermions and the squark production cross section is reduced due to the suppressed $t$-channel gluino contribution. 
\end{itemize}

The LHC phenomenology of this model should be compared to both models with right-handed neutrinos and to the MSSM. 
\begin{enumerate}
\item In models with right-handed neutrinos that address the origin and size of the neutrino masses, the SM singlets are only produced in EW processes via their mixing with the SM neutrinos. Due either to small mixing angles between the right-handed neutrinos and the SM neutrinos or to large right-handed neutrino masses, their production rates are greatly suppressed at the LHC. However, the bi$\nu$o can be produced in decays of colored particles in the model we consider. Hence this is a neutrino-mass model that can currently be probed at the LHC.  

\item In this model all supersymmetric particles need to be produced in sparticle--antisparticle pairs due to the $U(1)_R$ symmetry. (At 13~TeV LHC, the main $\tilde{q}\tilde{q}^\dagger$-production channel is gluon fusion.) Furthermore, some sparticle-production channels, \emph{e.g.} t-channel gluino exchange, are not present due again to the $U(1)_R$ symmetry. Hence it is expected that the constraints on this model are weaker than the ones on MSSM \cite{Kribs:2012gx}. Furthermore in this model the lightest neutralino, namely the bi$\nu$o, decays promptly and produces a combination of jets, leptons and missing energy.
\end{enumerate}

%%%%%%%%%%%%%%%%%%%%%%
\subsection{Expected signals and search strategies}
\label{sec:signals}

\begin{figure*}[t]
  \centering
  \begin{subfigure}[t]{0.5\textwidth}
  \includegraphics[width=\textwidth]{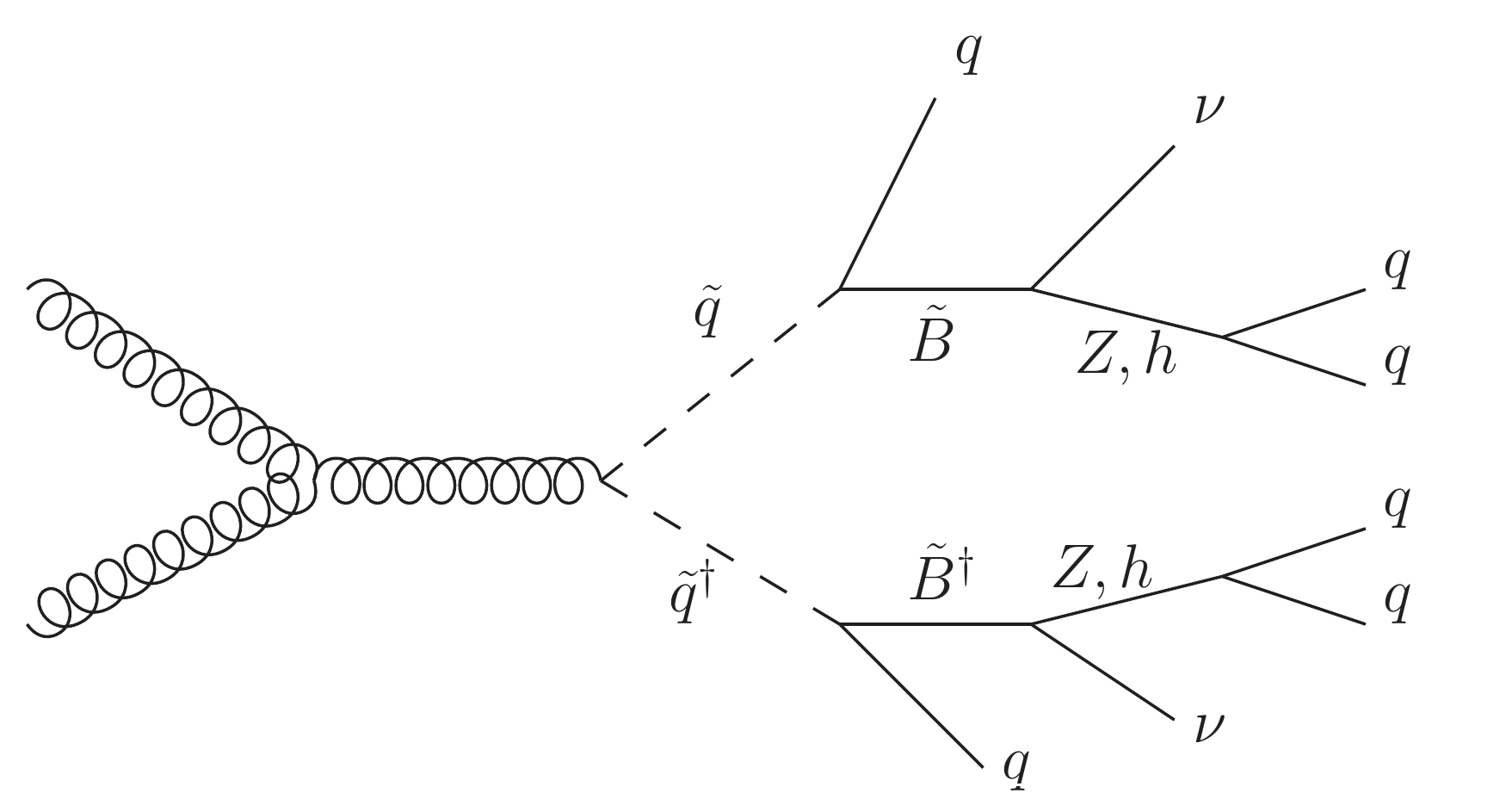}\label{fig:6jets}
  \caption{6 jets + missing energy}
  \end{subfigure}%
  ~
    \begin{subfigure}[t]{0.5\textwidth}
  \includegraphics[width=\textwidth]{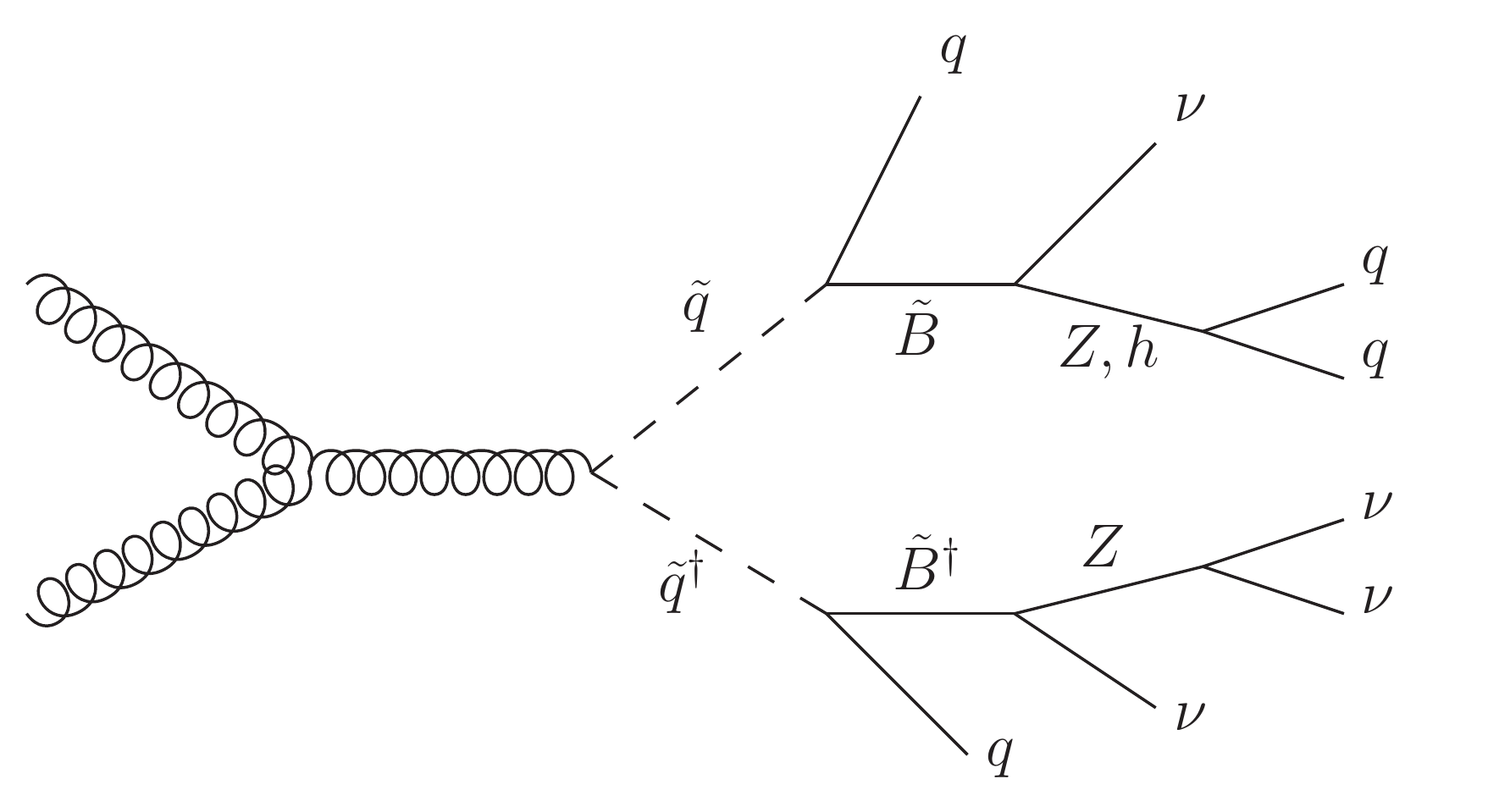}\label{fig:4jets}
  \caption{4 jets + missing energy}
  \end{subfigure}
   \caption{Final states with jets and missing energy. We recast current SUSY searches at ATLAS and CMS for this signal.} \label{fig:signals1}
\end{figure*}

\begin{figure*}[t]
  \centering
    \begin{subfigure}[b]{0.5\textwidth}
  \includegraphics[width=\textwidth]{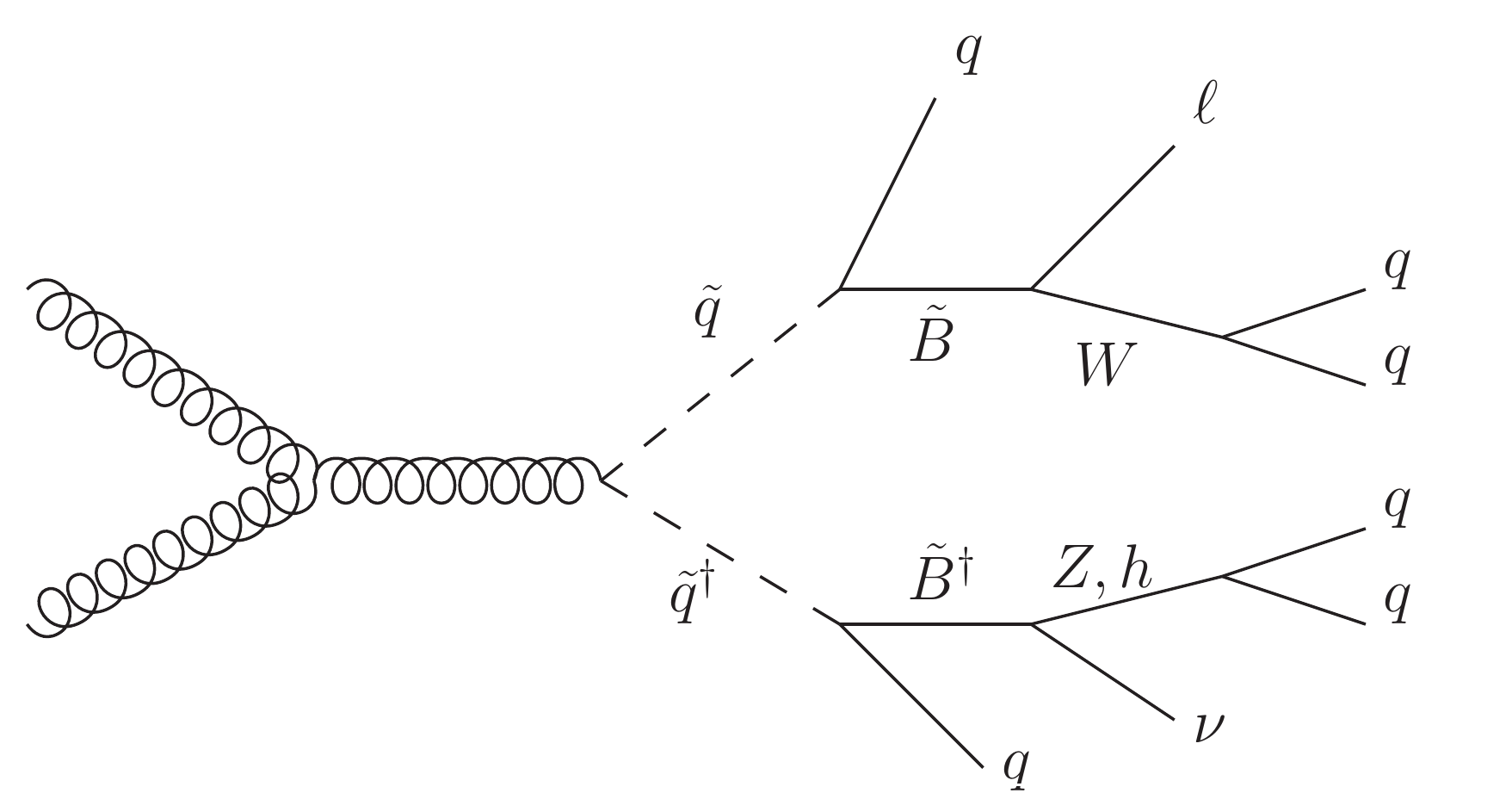}\label{fig:4jets}
  \caption{6 jets + 1 lepton + missing energy}
  \end{subfigure}%
    ~
  \begin{subfigure}[b]{0.5\textwidth}
 \includegraphics[width=\textwidth]{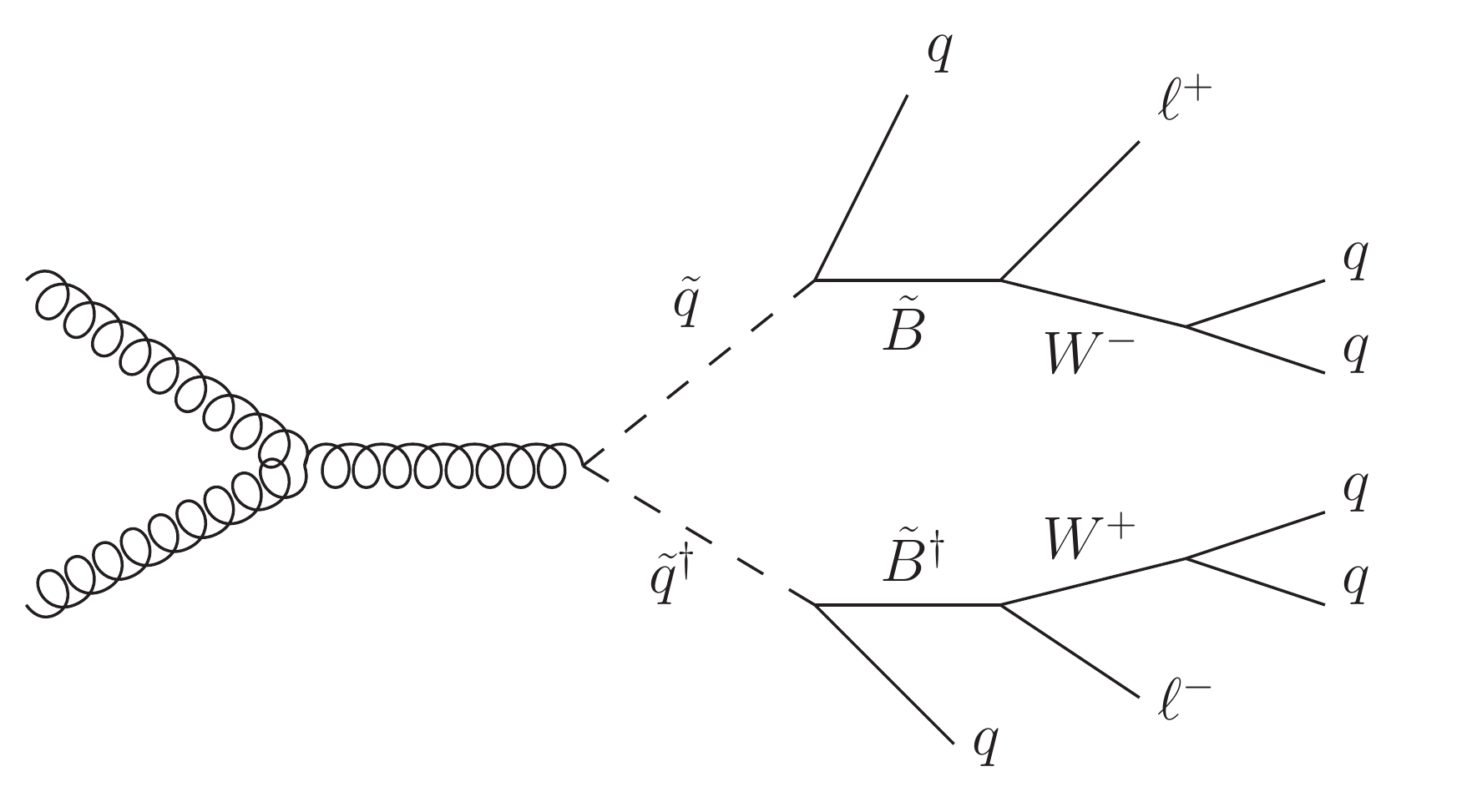}\label{fig:6jets2lep}
 \caption{6 jets + 2 leptons}
 \end{subfigure}
  \caption{Final states with leptons and missing energy. Leptoquark searches are recast for these signals.} \label{fig:signals2}
\end{figure*}

Due to the sparticle spectrum we assume, bi$\nu$os are predominantly produced via squark decays with $Br(\tilde{q}\to q \tilde{B}^\dagger)=1$. The bi$\nu$o subsequently decays through one of four possible modes: (i) $\tilde{B}\to\tilde{G}\gamma $; (ii) $\tilde{B}\to W^-\ell^+$; (iii) $\tilde{B}\to Z\bar{\nu}$; and 
(iv) $\tilde{B}\to h\bar{\nu}$. The first decay mode is strongly suppressed by the Planck mass, $\Gamma(\tilde{B}\to\tilde{G}\gamma)\sim \frac{M_{\tilde{B}}^5}{M_{\rm Pl}^2 m_{3/2}^2}\sim 10^{-8}$~eV. The rest of the decay modes are only suppressed by the neutrino-bi$\nu$o mixing angle and their branching ratios are approximately equal to 1/3. (Note that due to the $U(1)_{R-L}$ symmetry, $\tilde{B}\to W^+\ell^-$ decay is not allowed.)  The total decay width of the bi$\nu$o is $\Gamma_{tot}\sim M_{\tilde{B}}Y^2\sim M_{\tilde{B}}^3/\Lambda_M^2\sim O(10~{\rm MeV})$ for $M_{\tilde{B}}=500~$GeV and $\Lambda_M=100~$TeV. Hence, it decays promptly to final states, which include a combination of jets, leptons and missing energy. We show some of the final states with large branching fractions in Table \ref{table:BR} and Figures \ref{fig:signals1}-\ref{fig:signals2}.

\begin{table}[t]
\centering
\begin{tabular}{|c|c|c|}
\hline
Signal	&	Branching fraction	&	LHC searches \\
\hline\hline 
$6 j + \met$	&	20$\%$	&	ATLAS \cite{ATLAS-CONF-2017-022}, CMS \cite{Khachatryan:2016kdk, Khachatryan:2016epu, Sirunyan:2017cwe} \\
\hline
$6 j + 1\ell + \met$	&	15$\%$	&	ATLAS \cite{Aad:2016qqk, Aaboud:2016lwz}, CMS \cite{Khachatryan:2016epu, Khachatryan:2016iqn} \\
\hline
$4 j + 2\ell + \met$	&	6$\%$	&	ATLAS \cite{Aaboud:2016zpr, Aaboud:2016qeg}, CMS \cite{Khachatryan:2016iqn, Khachatryan:2017qgo} \\
\hline
$4 j + \met$	&	5$\%$	&	ATLAS \cite{ATLAS-CONF-2017-022}, CMS \cite{Khachatryan:2016kdk, Khachatryan:2016epu, Sirunyan:2017cwe} \\
\hline
$6 j + 2\ell$	&	3$\%$	&	ATLAS \cite{Aaboud:2016qeg}, CMS \cite{CMS-PAS-EXO-17-003}\\
\hline
\end{tabular}
\caption{Some of the signals that are produced by bi$\nu$o production and subsequent decays in the model described in Section~\ref{sec:model} with their branching fractions and relevant LHC searches. Here the leptons $\ell = e,\mu$ and $j=u,d,s,c$.} \label{table:BR}
\end{table}

We emphasize the importance of final states with leptons, \emph{e.g.} $6j+2\ell$ and $6j + 1\ell+\met$, as  smoking-gun signals in determining if bi$\nu$o is the source of neutrino mass generation, see Fig.\ref{fig:signals2}. The bi$\nu$o-neutrino mixing angle is $\theta_i \simeq \frac{Y_i v}{M_{\tilde{B}}}$ where $Y_i$ is given in (\ref{eq:YG}). The branching ratio of bi$\nu$o into different lepton species is fully determined by the neutrino mixing parameters. For example, in searches for first- and second-generation leptoquarks, relative rates of $ee:\mu\mu = 1:16$ and $e\nu:\mu\nu= 1:2$ are expected\footnote{Note that these branching fractions are given for the case where the phases in the PMNS matrix are set to zero. The matrix elements, hence the branching ratios, will change for non-zero  phases~\cite{Gavela:2009cd}.}. 

%%%%%%%%%
\subsection{Analysis}

Our model is implemented in \feynrules{} \cite{Alloul:2013bka} and the events are generated with \madgraph{5} \cite{Alwall:2014hca}, using \pythia{8} \cite{Sjostrand:2014zea} for parton shower and hadronization, and \delphes{} \cite{deFavereau:2013fsa} for detector simulation at $\sqrt{s}=13$ TeV and $\mathcal{L}=36~\text{fb}^{-1}$. We use the default settings for jets in \madgraph{5} with $R=0.4, ~p_{Tj}>20$ GeV and $|\eta_j|<5$. We generate signal events for bi$\nu$os in the mass range $100~{\rm GeV}<\mbino<\msq$ with a common squark mass for first and second generation squarks, $200~{\rm GeV}<\msq<1200~$GeV, in 50~GeV mass increments. We set all other sparticle masses to 10~TeV such that they are decoupled. As the bi$\nu$o mass gets closer to the squark mass, the computational time required to generate events increases. Hence, we do not consider splittings smaller than $25\GeV$, \iee\ $\msq-\mbino \ge 25~\GeV$. For $\mbino\lesssim 90~$GeV, the gauge bosons are off-shell and the phase space and the energy distribution of the final states are different. We leave a study of light bi$\nu$os to future work and focus on $\mbino>100~$GeV. 
 
We find that currently the most constraining search is the jets$ + \met$ final state due its large branching ratio and the integrated luminosity used in available analyses.  At the partonic level there are processes leading to 6q$ + \met$ and 4q$ + \met$ final states, see Figure~\ref{fig:signals2}.  We analyze this search in detail and use it to constrain the parameter space of the bi$\nu$o model.

We use the $m_{\rm eff}$-based  analysis given by ATLAS \cite{ATLAS-CONF-2017-022}. The observable $m_{\rm eff}$ is defined as the scalar sum of the transverse momenta of the leading jets and missing energy, $\met$. Taken together with $\met$, $m_{\rm eff}$ strongly suppresses the multijet background. There are 24 signal regions in this analysis. These regions are first divided according to jet multiplicities (2-6 jets).
 Signal regions with the same jet multiplicity are further divided according to the values of $m_{\rm eff}$ and the $\met/m_{\text{eff}}$ or $\met/\sqrt{H_T}$ thresholds. 
 In each signal region, different thresholds are applied on jet momenta and pseudorapidities to reduce the SM background.
Constraints on the smallest  azimuthal  separation  between $\met$ and 
 the momenta of any of the reconstructed jets further reduces the multi-jet background. Two of the signal regions require two large radius jets and in all  signal regions the required  jet momentum is $p_T>50~$GeV and missing energy $\met>250~$GeV. 
 The thresholds on the observables which characterize the signal regions have been chosen to target models with squark or gluino pair production and direct decay of squarks/gluinos or one-step decay of squark/gluino via an intermediate chargino or neutralino.

In order to identify the allowed parameter points we compare the  signal cross section to the measured cross section limits at  95$\%$ C.L. in all 24 signal regions using the code from \cite{Asadi:2017qon}. If the signal cross section of a parameter point exceeds the measured  cross section at 95$\%$ C.L. in at least one bin we take this parameter point to be ruled out. 
  
We also analyze the expected exclusion limits at the end of  LHC Run 3 with $\sqrt{s}=13~$TeV and $\mathcal{L}=300~\text{fb}^{-1}$, by rescaling with the luminosity the expected number of signal and background events, as given in~\cite{ATLAS-CONF-2017-022}. In order to obtain the allowed parameter region at a high-luminosity LHC we use the median expected exclusion significance \cite{Kumar:2015tna}
\begin{align}
Z_{exc}=\Big[2\left(s-b \log\left(\frac{b+s+x}{2b}\right)-\frac{b^2}{\Delta_b^2}\log\left(\frac{b-s+x}{2b}\right)\right)-(b+s-x)(1+\tfrac{b}{\Delta_b^2})
\Big]^{1/2}~,
\end{align}
with
\begin{align}
x=\big[(s+b)^2-4sb\tfrac{\Delta_b^2}{(b+\Delta_b^2)}\big]^{1/2}~,
\end{align}
where $s$ is the signal, $b$ is background and $\Delta_b$ is the uncertainty on the background prediction. For a 95$\%$ C.L. median exclusion, we require $Z_{exc}>1.645$.  We assume, as a conservative estimate, that the relative background uncertainty after $300~\text{fb}^{-1}$ remains the same as it is now, as presented in \cite{ATLAS-CONF-2017-022}.  The estimate that $\Delta_b/b$ is constant could be improved upon, especially if the background is estimated from data in sidebands.

%%%%%%%%%
\subsection{Results and discussion}
\label{sec:results}

\begin{figure}[t]
\centering
\includegraphics[scale=0.5]{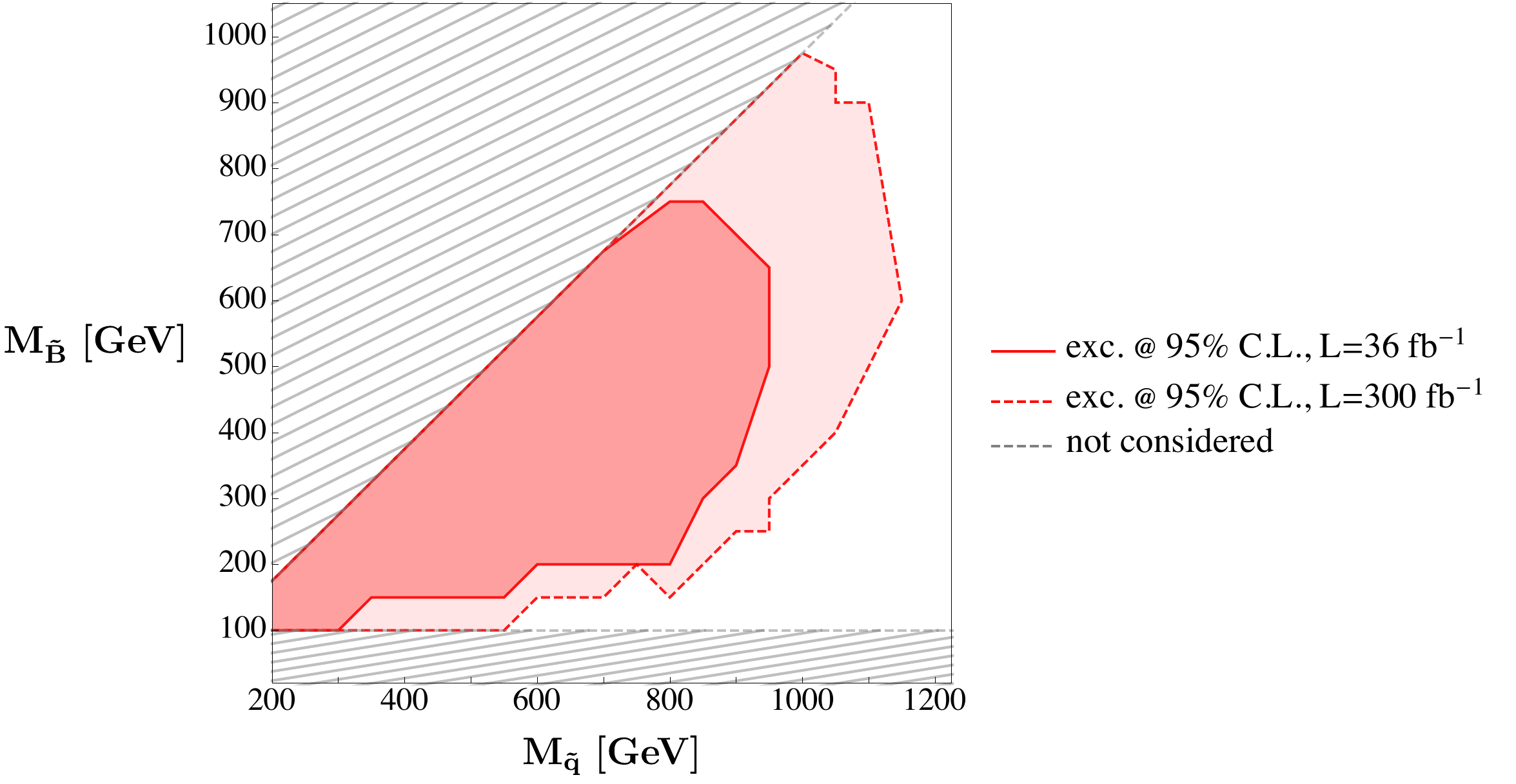}%\hspace{10mm}
\caption{\label{fig:Plotexcatlas} Current and forecasted 95\% exclusion limits from searches for  jets$+ \met$ final state in the squark mass ($\msq$) -- bi$\nu$o mass ($\mbino$) plane. The dark red region is excluded by our recast of the ATLAS analysis \cite{ATLAS-CONF-2017-022} which uses $\mathcal{L}=36~{\rm fb}^{-1}$ data at $\sqrt{s}=$~13~TeV. The red dashed line shows  a  forecast for $\mathcal{L}=300~\text{fb}^{-1}$ at $\sqrt{s}=$~13~TeV. We have not analyzed the parameter ranges in the gray striped regions, which correspond to regions where bi$\nu$o is heavier than squarks or the bi$\nu$o is lighter than the SM gauge bosons.}
\end{figure}

We show 95\% exclusion limits on squark and bi$\nu$o masses for current and forecasted searches in Figure~\ref{fig:Plotexcatlas}.
We find that squarks heavier than 950~GeV are not excluded for any bi$\nu$o mass by current LHC data with $\sqrt{s}=13~$TeV and $\mathcal{L}=36~{\rm fb}^{-1}$. In the mass regions we analyzed, bi$\nu$o masses 100--150~GeV are not currently excluded for squark masses above 350~GeV, as the resulting jet momenta and missing energy do not pass the search cuts. We also project limits for a high-luminosity LHC with $\sqrt{s}=13~$TeV and $\mathcal{L}=300~{\rm fb}^{-1}$. This forecast shows that even with this luminosity upgrade, as long as the same cuts are used in the analysis, the LHC can probe squark masses up to 1150~GeV. However, bi$\nu$o masses lighter than 150~GeV for $\msq>800$~GeV will still be allowed. 

We find that constraints in the parameter region we considered come from (4-6)-jet signal regions with small to medium values of $m_{\rm eff}$ for $\mathcal{L}=36~{\rm fb}^{-1}$. For our forecast to $\mathcal{L}=300~{\rm fb}^{-1}$, we find that constraints for $\msq<800$~GeV mostly come from a 4-jet region with small $m_{\rm eff}$ while a 6-jet region with medium  $m_{\rm eff}$ is most constraining for $\msq>800$~GeV.

In discussing our results we emphasize the differences between this model and some other ($R$-symmetric) SUSY models. 
\begin{enumerate}
\item In this model the gluinos are heavy and decoupled. Furthermore due to the $R$-symmetry some important squark production channels are not allowed. Hence, compared to the MSSM~\cite{SUSYxsec}, the squark--antisquark production cross-section is $O(0.1)$ smaller. 
\item Due to the sparticle spectrum we assume, squarks decay to a quark and the lightest neutralino 100\% of the time. The lightest neutralino, which we take to be purely bi$\nu$o, decays promptly to gauge bosons and leptons due to a broken $U(1)_{R-L}$ symmetry. In comparison to MSSM scenarios where the missing energy is carried by the neutralino, in this model there would be cascade decays and the missing energy is carried by light neutrinos. 
\item Similarly, due to the large number of jets and how the missing energy is distributed in this model, constraints on squark and bi$\nu$o masses are expected to be different than some other $R$-symmetric models, \emph{e.g.} \cite{Kribs:2012gx}. Although it is not straightforward to make a direct comparison, we point out that in \cite{Kribs:2012gx} the LSP is massless and the most constraining signal region contains only 2 jets whereas in this model the constraining signal regions contain 4-6 jets. The authors in \cite{Kribs:2012gx} mention as the LSP mass is increased to 300~GeV, all constraints disappear. However, note that, even with a finite mass, the LSP in that work does not decay.  We emphasize that we do not consider the region where $\mbino<100~$GeV. In this region the bi$\nu$o would decay via off-shell gauge or Higgs bosons. Due to the low mass of the bi$\nu$o, final states may not pass the missing energy and jet momentum cuts in the current analysis. We leave an analysis of this region to future work.

\item The closest study to ours is done in \cite{Frugiuele:2012kp}. In addition to some technical differences between the two models, in~\cite{Frugiuele:2012kp} the authors fix the lightest neutralino mass to be 1~TeV while we do a scan over both the squark and the bi$\nu$o masses. In \cite{Frugiuele:2012kp} the limit on the squark mass is found to be $\msq\simeq 650~$GeV by using an ATLAS jets+$\met$ analysis \cite{Aad:2012hm} at $\sqrt{s}=$~7~TeV with $\mathcal{L}=4.7~{\rm fb}^{-1}$ data. In our work we do not consider the region where $\mbino> \msq$. In this region the bi$\nu$o decays off-shell and it is expected that energy will be distributed to jets and missing energy democratically. 
 We expect the bound on the squark masses coming from the ATLAS anaylsis we use~\cite{ATLAS-CONF-2017-022} to be similar to that given by our most constraining signal region,  $5j+\met$, \iee~$\msq> 950$~GeV.
\end{enumerate}

We also analyzed the final state with $6 j + 2\ell$, which is a possible smoking gun signature for this model as the branching fractions of the bi$\nu$o to different lepton families is fully determined by the neutrino mixing parameters. We recast the CMS leptoquark analysis \cite{CMS-PAS-EXO-17-003}, which looks for a final state of two muons and two jets produced in the decay of a leptoquark pair. We find that this analysis currently has a very small exclusion power due to the small signal-to-background ratio ($S/B\sim 10^{-2}$). 

%%%%%%%%%%%%%%%%%%%%%%%%%%%%%%%%
\section{Conclusions}\label{sec:conclusions}

LHC constraints on sparticle masses in the MSSM are becoming more and more stringent.  Avoiding these strong experimental constraints and keeping superpartners light often leads to considering extensions of the MSSM.  These extensions are characterized either by adding additional operators (\egg~R-parity violation) or adding additional fields (\egg~Dirac gauginos).  We studied one such extension, with additional fields, which allows for a global $U(1)_{R-L}$ symmetry on the supersymmetric sector.   This leads to phenomenology associated with both R-parity violation and Dirac gauginos.  It was previously shown in \cite{Coloma:2016vod}  that the role of right-handed neutrinos can be played by one of these Dirac gauginos, the pseudo-Dirac bi$\nu$o, and that the observed neutrino mass spectrum can be achieved.  

We considered a scenario where the lightest neutralino is a pure bi$\nu$o, and this state is the lightest SM superpartner.  The squarks, which have a QCD production cross section, decay to the bi$\nu$o.  The mixing of this state with SM neutrinos means that it in turn can decay, despite the presence of a $U(1)_{R-L}$ symmetry.   The bi$\nu$o decays to a combination of quarks, leptons and missing energy.  We investigated the LHC constraints on this model and found the strongest comes from a recast of the most recent ATLAS analysis with $\sqrt{s}=13~$TeV and $\mathcal{L}=36~{\rm fb}^{-1}$, see Figure~\ref{fig:Plotexcatlas}. The constraints go up to only $\msq=950~$GeV and squarks as light as 350~GeV are allowed for $\mbino=100-150~$GeV. We also forecast constraints for $\mathcal{L}=300~{\rm fb}^{-1}$ at $\sqrt{s}=13~$TeV and show that high-luminosity LHC can probe up to $\msq=1150~$GeV if the same cuts for the jets+$\met$ analysis are used. However, even with the high-luminosity, low bi$\nu$o masses cannot be excluded. The flavor of the charged lepton in the bi$\nu$o decay depends upon the neutrino mixing parameters and thus the LHC is potentially sensitive to parameters in the neutrino sector, for instance through flavor ratios in leptoquark searches.

While our analysis indicates that, in these models, the squarks may be as light $950$ GeV for any bi$\nu$o mass, and as light as 350~GeV for bi$\nu$o between 100--150 GeV, it is intriguing to wonder if they can be even lighter.  We have not investigated the bounds for bi$\nu$o mass below $100$ GeV, nor the region with $M_{\tilde{B}}> M_{\tilde{q}}$.  It is also an interesting question to understand what are the ideal set of cuts for the jet+$\met$ final state to probe this model. Most importantly, the smoking-gun signals involving lepton final states need careful attention to find the best discovery path for this model. In a separate direction, the viability of gravitino/goldstino dark matter in this model requires detailed calculations of their production mechanisms given the sparticles masses allowed by LHC data.

\section*{Acknowledgements}

We thank Pilar Coloma for her collaboration in the early stages of this work. We are grateful to Angelo Monteux for sharing his analysis code with us as well as his help with running the code. SI acknowledges support from the University Office of the President via a UC Presidential Postdoctoral fellowship and partial support from NSF Grant No.~PHY-1620638. This work was performed in part at Aspen Center for Physics, which is supported by NSF grant PHY-1607611. JG has received funding/support from the European Union's Horizon 2020 research and innovation programme under the Marie Sklodowska-Curie grant agreement No 674896.
PF was supported by the DoE under contract number DE-SC0007859 and Fermilab, operated by Fermi Research Alliance, LLC under
contract number DE-AC02-07CH11359 with the United States Department of Energy.

\bibliographystyle{JHEP}

\providecommand{\href}[2]{#2}\begingroup\raggedright\endgroup

\end{document}